\documentstyle[emulateapj,epsf,amstex,astron_mlb]{article}
\lefthead{Balogh \& Morris}
\righthead{H$\alpha$ Photometry of Abell 2390}
\begin{document}
\newcommand{\oii}{[OII]$\lambda$3727}
\newcommand{\hd}{$W_{\circ}(H\delta)$}
\newcommand{\ebv}{$E_{\footnotesize \bv}$}
\newcommand{\br}{{\em B-R}}
\newcommand{\OII}{[O{\small\rm II}]}
\newcommand{\OIII}{O{\small\rm III}}
\newcommand{\NII}{N{\small\rm II}}
\newcommand{\SII}{S{\small\rm II}}
\newcommand{\CaII}{Ca{\small\rm II}}
\newcommand{\ArIII}{Ar{\small\rm III}}
\newcommand{\HI}{H{\small\rm I}}
\newcommand{\HII}{H{\small\rm II}}
\newcommand{\WOII}{$W_{\circ}$(O{\small\rm II})}
\newcommand{\wha}{$W_{\circ}(H\alpha)$}
\newcommand{\whd}{$W_{\circ}(H\delta)$}
\newcommand{\ow}{\WOII}
\submitted{MNRAS, Accepted July 3, 2000}
\title{H$\alpha$ Photometry of Abell 2390}

\author{Michael L. Balogh\altaffilmark{1, 2, 3},
Simon L. Morris\altaffilmark{2, 4}}
\altaffiltext{1}{\small{Department of Physics \& Astronomy, University of Victoria, Victoria, BC, V8X 4M6, Canada. }}

\altaffiltext{2}{\small{Visiting Astronomer, Canada--France--Hawaii Telescope, which is operated by the National Research Council of Canada, le Centre Nationale de la Recherche Scientifique, and the University of Hawaii.}}

\altaffiltext{3}{Present address: Department of Physics, University of Durham, South Road, Durham, England DH1 3LE. email: M.L.Balogh@@Durham.ac.uk}

\altaffiltext{4}{\small{Dominion Astrophysical Observatory, Herzberg Institute of Astrophysics, 
National Research Council, 5071 West Saanich Road, Victoria, B.C., Canada V8X 4M6. email: Simon.Morris@@hia.nrc.ca}}
\begin{abstract}
We present the results of a search for strong H$\alpha$ emission line galaxies
(rest frame equivalent widths greater than 50\AA) in the $z\approx0.23$ cluster
Abell 2390.  The survey contains 1189 galaxies over 270 \sq\arcmin, and
is 50\% complete at $M_r\approx-17.5+5\log{h}$.
The fraction of galaxies in which H$\alpha$ is detected at the 2$\sigma$ level rises
from 0.0 in the central regions (excluding the cD galaxy) to 12.5$\pm$8\% at $R_{200}$.
For 165 of the galaxies in our catalogue, we compare the H$\alpha$ equivalent widths with
their \OII$\lambda3727$ equivalent widths, from the CNOC1 spectra.  The fraction of strong H$\alpha$ emission 
line galaxies is consistent with the fraction 
of strong \OII\ emission galaxies in the CNOC1 sample:
only $2\pm1$\% have no detectable \OII\ emission and yet 
significant ($>$2$\sigma$) H$\alpha$ equivalent widths.  Dust obscuration, non-thermal
ionization, and aperture effects are all likely to contribute to this
non-correspondence of emission lines.
We identify six spectroscopically 'secure' k+a galaxies (\ow$<5$\AA\ and \hd$\gtrsim5$\AA);
at least two of these show strong signs in H$\alpha$ of star formation in regions that are covered by
the slit from which the spectra were obtained.  Thus, some fraction of
galaxies classified k+a based on spectra shortward of 6000 \AA\ are likely to be undergoing
significant star formation.   
These results are consistent with a 'strangulation' model for cluster galaxy
evolution, in which star formation in cluster galaxies is gradually decreased, and is
neither enhanced nor abruptly terminated by the cluster environment.

\end{abstract}


\section{Introduction} \label{sec-intro}
The study of large scale gradients in $z\lesssim 0.5$ clusters has benefited lately from
a wealth of observational data.  Radial gradients in stellar populations
and star formation rates (SFRs) have been measured out to beyond the virial radius in fifteen
moderate redshift clusters in the Canadian Network for Observational Cosmology (CNOC1) 
sample (\cite{A2390,1621,B+97};\cite*{B+98,PSG});
the MORPHS collaboration (\cite{Smail}) have been able to do similar work at smaller radii (\cite{D+99,P+99}), 
with the added advantage of {\it Hubble Space Telescope} (HST) imaging to study galaxy morphologies in these
clusters (\cite{C+94,D+97,C+98}).  These studies, and others,  have shown that there are strong gradients
in the mean SFR;
recent modelling by Balogh et al. (\cite*{infall}) 
suggests that it is reasonable to interpret this as a consequence of 
gradients in typical mass accretion times.

There is an intriguing class of unusual galaxies with strong $H\delta$ absorption, but no observable \OII$\lambda$3727 
emission (e.g. \cite{DG83,CS87,PSG}) which have been an important focus of the above studies.
The dust--free models constructed to reproduce these strong Balmer absorption lines have two important
characteristics: (1) active star formation must have ceased fairly abruptly, within $\lesssim 1$ Gyr;
and (2) the galaxies with the strongest H$\delta$ lines require this abrupt truncation
to be immediately preceded by a strong {\it increase} in the SFR.
The presence of these galaxies in some clusters has therefore led to the suggestion that
star formation is abruptly quenched in a large fraction of infalling cluster
galaxies (e.g. \cite{Barger}).  However, there
is mounting evidence that some of these unusual spectra may result from 
patchy dust obscuration (\cite{P+99,Smail-radio,PW}).  In this case, the galaxies
are actually still
undergoing strong star formation, but the most massive stars are heavily obscured.
Longer lived A-stars migrate out of these dusty regions and become a more widespread
population, dominating the continuum light even though star formation
is still present.
In light of this, we want to move away from empirical definitions,
which are based on the strengths of spectral features, and propose four quantitative, physical
definitions which we will use throughout this work.  
\begin{itemize}
\item {\bfseries Truncation: } A decline in SFR from a significant rate (i.e., one that can
account for the formation of a substantial fraction of the galaxy's stars over a Hubble time, say
$\dot{M}>0.1$M$_\odot$yr$^{-1}$)
to negligible or undetectable levels in $\lesssim$1 Gyr.
\item {\bfseries Starburst: } An episodic increase in the SFR, during which the galaxy increases
its stellar mass by more than 10\%, in $\lesssim 1$ Gyr.  Spectrophotometric modelling suggests that
a burst strength of at least this size is necessary to produce the galaxy spectra with the
strongest Balmer absorption lines.
\item {\bfseries Post-starburst: } The phase lasting $\sim 1$ Gyr after the truncation of a starburst. 
It is during this time that Balmer lines are significantly enhanced, relative to galaxies undergoing
quiescent star formation.
\item {\bfseries Strangulation: } A gradual decline in the SFR of a galaxy,
due to the absence of halo gas that is required to continually fuel the disk of normal star
forming galaxies.  It has been suggested that this is the primary mechanism responsible for
the differences between field and cluster galaxies (\cite{infall}).
\end{itemize}

If dust-obscured star formation is occurring in H$\delta$-strong galaxies, 
then most of the evidence for truncated star formation disappears.  
Instead, the systematically
lower SFRs of cluster galaxies relative to the field may be due to the more gradual
'strangulation' process, since galaxies are unlikely to be able to retain a gaseous
halo in the cluster environment (\cite{LTC,infall}).  
Moss \& Whittle (\cite*{MW,MW00}) and Moss et al. (\cite*{MWP98}) have demonstrated that 
spiral galaxies in rich clusters show enhanced, circumnuclear H$\alpha$ emission, relative
to the field, which they suggest are due to tidal effects.  However, it is not clear
if the frequency and intensity of this activity is sufficient to be responsible for
the transformation of the field spiral population into the dominant S0 population in
clusters.

In many spectroscopic studies of high redshift galaxies, including the CNOC1 and MORPHS
surveys discussed above, slit spectroscopy of
the \OII\ line is used as evidence for star formation.  
Kennicutt (\cite*{K92}) has shown
that the strength of this emission line correlates well with SFR.
However, there are several problems inherent in its use for studies of this kind.
In particular:
\begin{itemize}
\item {\bfseries Physical Interpretation: }The strength of the \OII\ line is sensitive to not only
the ionizing flux from massive stars, but also to the metallicity and ionization state
of the emitting gas.  Primarily because of variations in these quantities, the correlation
between \OII\ strength and SFR shows considerable scatter, and varies considerably between
samples (e.g., \cite{G+89,K92,T+99}).  However, the presence of \OII\ emission is still
a reliable indication that star formation is present (except in the case of active galactic
nuclei), even though the absolute rate may be 
uncertain.

\item {\bfseries Dust Extinction: }The \OII\ line is a blue feature, and quite sensitive
to foreground dust extinction.  Although line equivalent widths 
are unaffected by uniform dust obscuration,
the derivation of SFRs requires estimating (or measuring) the line flux, which can be
significantly reduced by dust extinction.  Weaker lines will become undetectable in the
presence of dust, and the number of star forming galaxies will be underestimated.
Furthermore, if dust obscuration is patchy, the line emission may be more extincted
that the continuum light, which will serve to reduce the line equivalent width,
as well as the flux (e.g., \cite{Calzetti}). 

\item {\bfseries Slit Sampling: }In most spectroscopic studies of galaxies at $z\approx0.3$,
the spectra are obtained from a narrow ($\lesssim 2$\arcsec) slit centred on the peak of the continuum light, i.e., the
centre of the galaxy.  While this can cover a fairly large region at moderate redshifts, such
observations may still exclude much of the flux from an extended disk component, where copious amounts
of star formation often take place.  Thus, measurements from slit spectroscopy do not always
correspond directly to the total, luminosity-weighted average properties of the galaxy as a whole.
\end{itemize}

One can move a long way toward overcoming the above difficulties by undertaking imaging
surveys in H$\alpha$ light; this emission line is a more reliable star formation
indicator than \OII, as it is less sensitive to dust extinction and metallicity.  Underlying
stellar absorption is generally less than 5\AA\ (\cite{K92}).  
Assuming a Salpeter initial mass function (IMF) for stellar
masses ranging from 0.1 to 100 $M_\odot$, Kennicutt et al. (\cite*{K+94}) find the
following relationship between H$\alpha$ luminosity, L(H$\alpha$) and SFR
(for solar metallicity):
\begin{equation}\label{eqn-sfr}
\mbox{SFR}(M_\odot \mbox{yr}^{-1})=7.9\times 10^{-42}{L(H\alpha) \over \mbox{ergs s}^{-1}}.
\end{equation}
As recently reviewed by Kennicutt (\cite*{Kenn_review}), other calibrations have been
published, with a variation of about 30\% arising mostly from differences in the assumed
initial mass function, but also, to some extent, from the nature of the models used.  
In particular, the coefficient in Equation \ref{eqn-sfr} is 13\% smaller than 
that determined by Kennicutt (\cite*{K92}, as used in Balogh et al. \cite*{B+97})
and 50\% smaller than that determined by Barbaro and Poggianti (\cite*{BP}, adopted 
in Balogh et al. \cite*{B+98}).
Although the H$\alpha$ line is much less sensitive to extinction than \OII, this
effect is still the most important source of systematic error (\cite{Kenn_review}).  Typical estimates
of the mean extinction for nearby spirals are A(H$\alpha$)=0.5--1.8 mag (\cite{K83,N+97,CD86,K+87,vdH+88,C+96}); furthermore
A(H$\alpha$) is certainly not the same for all galaxies, and it is likely correlated with galaxy
type and SFR.  In particular, the extinction could be considerably higher in starburst
galaxies which are targeted by the present survey. 

In this work, we present the results of narrow band H$\alpha$ imaging of galaxies in the CNOC1 cluster
Abell 2390, an object which has been well studied previously by Abraham et al. (\cite*{A2390}).
We use this data to determine:
\begin{itemize}
\item[1.]  The fraction of cluster galaxies undergoing strong star formation ($\dot{M}\gtrsim 1 M_\odot $yr$^{-1}$).
We note that this alone cannot be used to determine the number of cluster-induced starbursts without similar 
measurements of the field galaxy population and a model of cluster infall.  A first attempt at such
modeling, based on \OII, is presented in Balogh et al. (\cite*{infall}).
\item[2.] The fraction of cluster galaxies which are undetected in \OII, but show H$ \alpha$ emission.
This allows us to quantify how much star formation is missed in spectroscopic surveys based on \OII,
and to investigate possible reasons for this omission.
\item[3.] The presence and nature of H$\alpha$ emission in the small k+a galaxy population.  Their detection
in H$\alpha$ would imply that these systems do not satisfy our definition of 
physical post-starburst galaxies.
\end{itemize}

In \S\ref{sec-obs} we describe the details of the observations made in 1998 at
the Canada--France--Hawaii Telescope (CFHT).  The reduction of this data and the details of the photometry,
are discussed in \S\ref{sec-data}.  Our catalogue is presented in \S\ref{sec-cat}, 
the results are shown in \S\ref{sec-results},
and our findings are summarised in the final section, \S\ref{sec-concs}.

\section{Observations} \label{sec-obs}
The data were obtained 
with the {\em OSIS} (Optical Sub--arc second Imaging Spectrograph) instrument in 
imaging mode, over four half nights at CFHT from June 21--24, 1998.  
Twenty--one (19 with acceptable transmission) pointings
were made, for a total areal coverage of about 270 \sq\arcmin.  
Only the central and extreme east/west portions of the CNOC1 strip
were observed, due primarily to a night lost to cloudy conditions.
For each pointing, we obtained images through three narrow-band filters (the on-line, a red continuum (RC)
and a blue continuum (BC)) and one broad band R filter.  The three narrow--band 
filters were chosen to isolate the H$\alpha$ emission line at the mean cluster redshift of $z=0.228$,
$\lambda=8059$ \AA.  An appropriate on--line filter was manufactured by Barr Associates, with
$\lambda_\circ=8071.5$ \AA, $\Delta\lambda=348$ \AA\ FWHM (398\AA\ between 
10\% transmission levels) and a peak transmission
of $\sim$90\%. Adequate filters
for the blue and red continuum observations were available
from CFHT\footnote{\#4701: $\lambda_\circ=7425$\AA, $\Delta \lambda_{\rm FWHM}=125$\AA\  and \#1814: $\lambda_\circ=8752$\AA, $\Delta \lambda_{\rm FWHM}=181$\AA.}.  These were chosen to encompass as much continuum as possible,
while avoiding H$\alpha$ emission from cluster members, and other strong emission
lines.  Using two continuum filters allows the determination of the continuum level
at H$\alpha$, even in the presence of a strong slope.
Although relatively isolated, the on--line filter will still sometimes be contaminated
with [\SII]$\lambda\lambda 6717,6731$ emission, in addition to  [\NII], while 
the red continuum filter is susceptible to contamination by the very weak [\ArIII]$\lambda 7136$ line.
Corrections are only made for 
[\NII]$\lambda\lambda6548,6583$ emission, using the relation [\NII]/H$\alpha$ $\approx$ 0.3
appropriate for strong emission line galaxies (\cite{T+99}).

Exposure times were 90 s for the R--band filter, 600 s for the on-line filter,
and 300s for each of the two continuum filters.  
A  summary of the
fields observed is presented in Table \ref{tab-log}.  
\begin{deluxetable}{cccc}
\tablewidth{0pt}
\footnotesize
\tablecaption{List of Observations\label{tab-log}}
\tablehead{\colhead{Date (1998)}&\colhead{Field}&\colhead{$\Delta$ RA ($\arcsec$)}&\colhead{$\Delta$ Dec ($\arcsec$)}}
\startdata
 21/6&  cc  &    0 &   0 \nl
 ``    &  e1c &  225 &   0 \nl
 ``   &  e1n &  225 & 170 \nl
 ``  &  cn  &    0 & 170 \nl
 23/6&  w1s & -225 & -170\nl
 ``   &  cs  &    0 & -170\nl
 ``    &  e1s &  225 & -170\nl
 ``    &  w4n & -900 &   90\nl
 ``    &  w3n & -675 &   90\nl
 ``    &  w2n & -450 &   90\nl
 ``    &  w2s & -450 &  -90\nl
 24/6&  w3s & -675 &  -90\nl
 ``    &  w4s & -900 &  -90\nl
 ``   &  e4n &  900 &   90\nl
 ``    &  e4s &  900 &  -90\nl
 ``    &  e3s &  675 &  -90\nl
 ``    &  e3n &  675 &   90\nl
 ``    &  e2n &  450 &   90\nl
\enddata
\tablecomments{Positions are measured relative to 21:53:39.2 +17:41:16 (J2000).}
\end{deluxetable}

\section{Data Reduction and Photometry} \label{sec-data}
Each image was reduced using standard procedures with the IRAF\footnote{IRAF is distributed by the National Optical Astronomy Observatories which is 
operated by AURA 
Inc. under contract with NSF.} {\it ccdproc} package.  The continuum and on-line images were aligned by matching
the centroids of several stars, and the astrometric solution, used for matching galaxies with
the CNOC1 sample, was computed from
stars in the U.S. Naval Observatory (USNO) A2.0 catalogue\footnote{The USNO SA2.0 catalogue is a product of the USNO Flagstaff Station and
is distributed by the Astrometry Department.}.  
Object detection, photometry and star/galaxy separation was done with the SExtractor
photometry package (\cite{sextractor}).  We estimated the background using a global background
map; this is inadequate for objects found in the haloes of bright stars
or near bright ghosts or reflections, and these 35 objects are excluded from the catalogues.
Objects are detected as 5 contiguous pixels more than 1$\sigma$ over the background, following
convolution with a Gaussian (5 pixels FWHM).   These detections are then passed through
a deblending algorithm, which rethresholds each object at 32 exponentially spaced levels,
and identifies peaks that contribute at least $10^{-6}$ of the flux of the blended
structure as distinct objects.  Due to the sensitivity of measured colours to this
deblending procedure, these measurements must be treated with some caution.  For this reason, we will
plot all deblended galaxies as separate symbols in the figures of this paper.  

Sources were detected separately on all four images for each field; these detections were then
matched, so the final catalogue contains only objects detected on all four images.  This
allows easy rejection of false detections due to noise spikes or residual cosmic rays.
Magnitudes in each filter were determined from the FLUX\_BEST quantity 
calculated by SExtractor; this uses an adaptive aperture approach
to estimate the total light associated with each object.  This approach is taken 
to allow for the fact that the PSF can vary significantly
between different images of the same field (in different filters).  We have verified that 
on-line/continuum ratios, measured in large enough apertures ($\gtrsim$ 2\arcsec\ diameter, using
the IRAF {\it phot} task) are consistent
with the ``total'' ratios measured in this manner, within the uncertainties.
Additional confidence that the equivalent widths measured in this way do not include additional,
significant systematic uncertainty is given by comparing independent measurements of the same
galaxy; this is discussed in detail below.

SExtractor star--galaxy classification is used to remove stellar objects from the final catalogue.
The classification
of bright objects ($r>21.7$) is at least reliable at the 95\% level, though there is some tendency to
classify compact members of deblended galaxy-galaxy pairs as stars.  SExtractor assigns a
``stellarity index'' between 0 and 1 to each galaxy, where the larger number indicates increased
likelihood that the object is a star.  By inspecting objects on several of our OSIS images,
we determined that almost all objects with stellarity index greater than 0.97 are stars, and we adopt this as
our threshold.  

Photometric zero points were determined from observations of a 
spectrophotometric standard star, BD+28$^\circ$ 4211 (\cite{O90}), in each filter. 
The relative throughput as a function of wavelength for the  filter and  detector combination
is required to determine the expected flux of this standard star in each of our filters.  This response
was determined by taking a dispersed, long slit image of
the flat field lamp through each filter, and dividing this by the spectrum obtained
without a filter in place.  We renormalise this response function
 to reach a maximum value of unity,
so that the value in each wavelength interval represents the relative contribution of flux in that interval,
accounting for both filter and detector response. 

\begin{figure*}
\begin{center}
\leavevmode \epsfysize=8cm \epsfbox{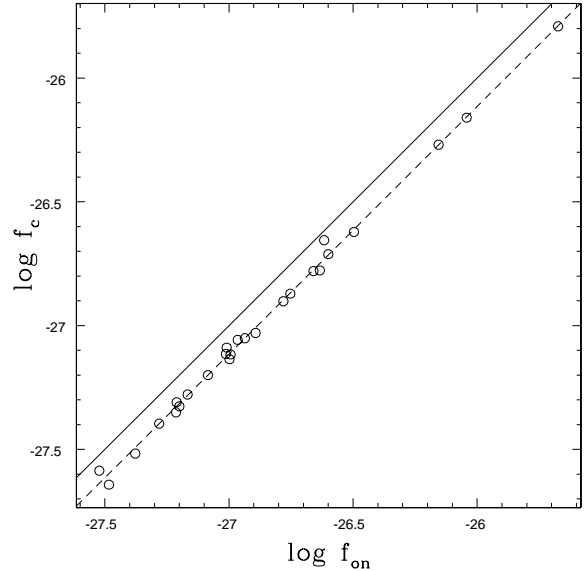}
\end{center}
\caption
{The logarithm of the flux in the averaged continuum ($f_c$) is plotted against 
the logarithm of the flux in the on--line filter, for
stars in the central field.  Error bars are omitted, as they are generally smaller
than the plotted symbols.  The two fluxes should be equal (as indicated by the {\it solid line}), since stellar
objects have no strong features at this wavelength.  Due to non-photometric conditions, the
continuum flux in this case is a factor of 1.3 fainter than the on-line flux ({\it dashed line}),
with an {\it r.m.s.} dispersion of about 5\% in this factor.
Thus, we divide the on--line flux by 1.3 for  all galaxies in this
field when equivalent widths are measured.  Similar calculations are done for the other fields.
\label{fig-ccstars}}
\end{figure*}

Since the nights were not photometric, the zeropoints are not
reliable to within more than $\sim$0.3 mag.  However, H$\alpha$ equivalent widths (\wha) can be 
more precisely determined, since they only depend on the relative flux between the on-line
and continuum images.  We calibrate this by assuming that
stellar objects have no significant absorption or emission at $\lambda\approx 8100$\AA, corresponding to
the wavelength of our on--line filter; thus, an appropriate scaling can be found such that
the mean equivalent width of stellar objects is equal to zero.
We first calculated the continuum flux of each galaxy by averaging its flux (in ergs s$^{-1}$  Hz$^{-1}$ cm$^{-2}$) 
in each of the two continuum
filters, RC and BC, to obtain $f_c$; this is appropriate because the central wavelengths of 
RC and BC are almost equally spaced on either side of the on--line filter in frequency.  We then selected a sample of 
unsaturated, unblended stellar
objects (with stellarity index $>$0.97, and free from image edges or bad columns), and measured
the ratio of the flux in the on--line filter, $f_{on}$, to $f_c$.  An example of this relation is shown in Figure
\ref{fig-ccstars}, where we compare these two fluxes for stars in the central field.  The tight correlation
suggests that it is fair to assume all stars in the field have an approximately featureless spectrum in this wavelength
range.
We then divided $f_{on}$ by the mean value of this
ratio ($s=1.30$ in this case), to obtain a scaled image $f_{on}^{\prime}$ for all galaxies in the field,
ensuring that the mean of $f_{on}^{\prime}-f_c$ is zero for stellar objects.  
For a given image, the {\it r.m.s.} of the scale factor $s$ is about $\Delta s / s\approx 0.05$, and
we adopt this as the relative uncertainty in this scaling. 

The rest frame equivalent width of the H$\alpha$ line for cluster members is given by 
\begin{equation}
W_{\circ}(H\alpha)=\Delta \lambda {f_{on}/s-f_c \over f_c}=\Delta \lambda {f_{on}^{\prime}-f_c \over f_c},
\end{equation}
where $\Delta \lambda$ is the rest frame width of the on--line filter, in \AA, which we take to be the FWHM of the
filter response function, 348 \AA.   In fact, the on-line filter also covers the adjacent [\NII] emission lines; 
all values of \wha\ presented here include contribution from these lines, and a correction will only
be made when SFRs are derived.  For galaxies with $z<0.2033$ or $z>0.2564$ this index will not measure 
the strength of H$\alpha$, though we retain this notation since cluster membership has not been determined
for all galaxies.
The uncertainty in \wha\ is computed using standard
independent error propagation, including the 5\% uncertainty in the scale factor $s$.  Note that, for large
values of $W_{\circ}(H\alpha)$, the relative
uncertainty in this quantity is larger than the relative uncertainty in the H$\alpha$ flux, $f_{on}^{\prime}-f_c$,
but always by less than 10\%.

\begin{figure*}
\begin{center}
\leavevmode \epsfysize=8cm \epsfbox{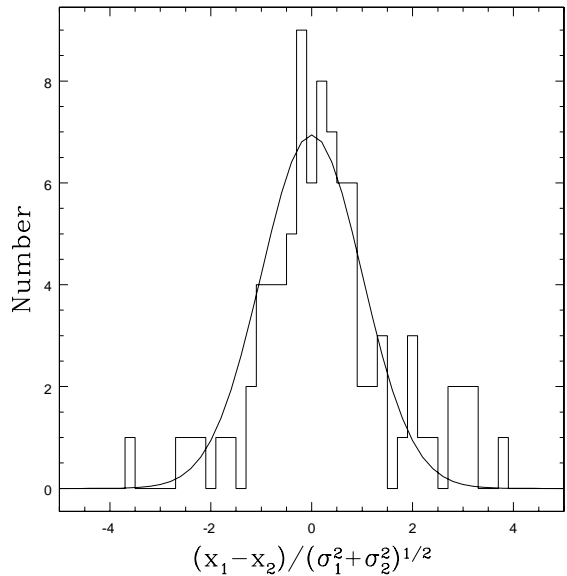}
\end{center}
\caption
{For multiply observed galaxies in the catalogue, we plot the ratio of the difference
between two independent \wha\ measurements ($x_1$ and $x_2$) and the quadrature sum of their associated errors.
A K--S test cannot distinguish between this distribution and that of a 
Gaussian with a mean of zero and variance of unity, shown by the
{\it solid line}.  This implies that the error estimates accurately reflect the reproducibility
of a measurement.
\label{fig-EW_err}}
\end{figure*}

Since adjacent OSIS fields overlapped by up to 30\arcsec, some galaxies appear on more than one image;
from these duplicates, only the best quality image was used.  We can use the duplicate observations to estimate
the reliability of our equivalent width measurements, by comparing the
difference between two measurements ($x_1-x_2$) with the quadrature sum of their uncertainties,
$\sqrt{\sigma_1^2+\sigma_2^2}$.  
In Figure \ref{fig-EW_err} we plot the distribution of the ratio of these two numbers, 
$\epsilon=(x_1-x_2)/\sqrt{\sigma_1^2+\sigma_2^2}$, excluding galaxies near image boundaries, and those
which were originally blended with another object.
If the true errors are Gaussian distributed with a variance given by our error
estimate, the distribution of $\epsilon$ should be  Gaussian with a mean of zero and variance of unity; this
is shown as the solid curve for reference.  A K--S test cannot significantly distinguish between this Gaussian and the
$\epsilon$ distribution (the probability that the two are drawn from the same distribution is 0.14).  
This implies that our error estimates reliably represent the reproducibility of
the equivalent width measurements, and that there are no large systematic effects.

\section{The Final Catalogue}\label{sec-cat}
Galaxies lying near the single bad column of the STIS2 chip, in the haloes of bright stars, or with
corrupted photometry (due to nearness to an image boundary or saturated pixels) were removed from the sample.
The final catalogue contains flux ratios (corresponding to \wha\ for cluster members) for 1189 galaxies, 
and will be made available
at the Astronomical Data Center (http://adc.gsfc.nasa.gov/adc/); the data can also be obtained directly
from M. Balogh.  A sample
entry in this catalogue is shown in Table \ref{tab-cat}.  The name of the field in which the object lies is given
in column 1, and the position on the CCD, in pixel coordinates, is given in columns 2 and 3.   
The J2000 coordinates of each object are shown in columns 4 and 5.
We take the position of the brightest cluster galaxy (BCG) to be the centre of the 
cluster, relative to which galaxy
positions are measured; offsets from this position, in arcseconds, are given in columns 6 and 7. 
From Carlberg et al. (\cite*{CNOC1}), $R_{200}$ for this cluster is 1.51h$^{-1}$
Mpc, 
where $R_{200}$ is the radius at which the mean interior mass density is equal to
200 times the critical density, and within which it is expected that the galaxies are
in virial equilibrium (\cite{GG,CER}).  This radius
corresponds to 624\arcsec\footnote{We adopt $\Omega_\circ=0.2, \Lambda=0$ for all cosmological dependent
calculations.}, and we normalise all projected cluster--centric distances ($R_{proj}$) to this value.
The flux (column 11) and equivalent width (column 13) of the feature at 8059\AA\ corresponds to H$\alpha$
for galaxies within $\sim 5\sigma$ of the cluster redshift;
the former quantity should only be used in a relative sense, due to systematic uncertainty in the zero point.  In column
15 we list the SExtractor photometry flags which indicate the presence of bright neighbours (1), a deblended
object (2), both (3) or neither (0).
\begin{deluxetable}{cccccccccc}
\tablewidth{0pt}
\footnotesize
\tablecaption{Sample Entry in H$\alpha$ Catalogue\label{tab-cat}}
\tablehead{
\colhead{(1)}&\colhead{(2)}&\colhead{(3)}&\colhead{(4)}&\colhead{(5)}&\colhead{(6)}&\colhead{(7)}&\colhead{(8)}&\colhead{(9)}\nl
\colhead{Field}&\colhead{x}&\colhead{y}&\colhead{RA}&\colhead{Dec}&\colhead{$\Delta$RA}&\colhead{$\Delta$ Dec}&\colhead{$r$}&\colhead{$\Delta r$}
}
\startdata
cc & 784.4 & 473.6 & 21:53:36.80 & 17:41:43.8 & 0. & 0. & 17.209 &0.004\nl
\hline\nl
\hline
(10) & (11) & (12) & (13) & (14) & (15) & (16) & (17) & (18)\\
$r_{\rm corr}$&flux &$\Delta$ flux&$W_\circ(H\alpha)$&$\Delta W_\circ(H\alpha)$&Flag&$R_{\rm weight}$&ppp&C.M.\\
\hline\\
17.258& 0.117 & 0.029 & 86.6 & 21.9 & 2 & 1.12 & 101084 & yes\nl
\enddata
\tablecomments{Column definitions:(1) Field name; (2-3) chip coordinates (pixels); (4-5) J2000 coordinates;
(6-7) Distance from central galaxy (arcseconds); (8-9) $r$ magnitude and uncertainty; (10) $r$ magnitude
corrected for zero-point based on CNOC1 catalogue; (11-12) H$\alpha$ flux in mJy,
and its uncertainty;
(13-14) Rest frame equivalent width of H$\alpha$
and its uncertainty (\AA); (15) SExtractor photometry flag in the on-line image; (16) incompleteness correction; (17) CNOC1 ppp number, if a spectrum is available;
(18) cluster membership as determined from the CNOC1 redshift.
The full catalogue will be available at the Astronomical Data Center.}
\end{deluxetable}

\subsection{Completeness}\label{sec-complete}
\begin{figure*}
\begin{center}
\leavevmode \epsfysize=8cm \epsfbox{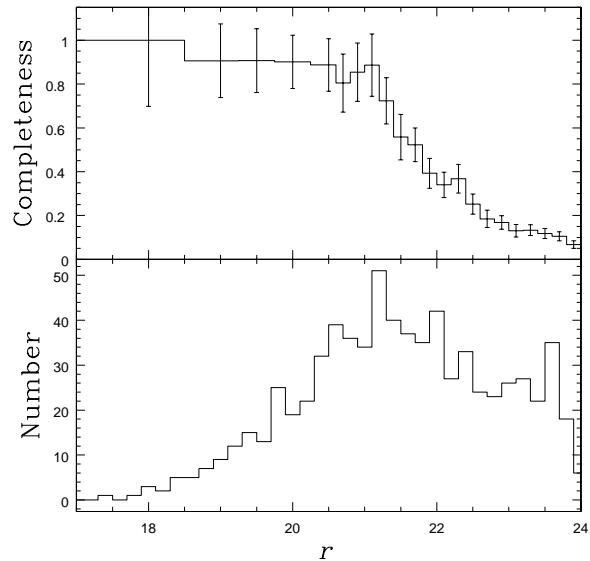}
\end{center}
\caption
{The {\it bottom panel} shows the total number of galaxies in the H$\alpha$ sample that correspond
to a galaxy in the CNOC1 sample, as a function of
$r$ (from the CNOC1 photometry).  In the
{\it top panel}, we plot the ratio of the number of galaxies in our final  
H$\alpha$ sample to the number of galaxies in
the CNOC1 photometric catalogue,
matched to identical areal coverage.  Error bars are 1$\sigma$, and are determined
as the value in each bin divided by $\sqrt{N}$, where $N$ is the number of galaxies  
from the H$\alpha$ sample in that bin.  The completeness drops
to 50\% at $r=21.7$.
\label{fig-complete}}
\end{figure*}

To estimate the completeness of our final catalogue, we compare our galaxy sample with the 
photometric CNOC1 galaxy sample, which is complete to about $r=23.5$ (\cite{A2390}).  Galaxies were matched in 
right ascension and declination coordinates, verified by interactive inspection in ambiguous
cases. 
The number of galaxies in our catalogue matched in this manner is shown in the bottom
panel of Figure \ref{fig-complete}, in bins of CNOC1 $r$ magnitude.  The 
difference between this distribution and that of
galaxies in the CNOC1 sample that lie within the same area of sky is shown in the top panel.
Note that this procedure does not account for the small number of galaxies in the present sample which have
no match in CNOC1; these usually arise from cases where our superior sampling and seeing is able to resolve
an object into multiple components.  There are also cases of discrepancy where objects are classified as stellar in
one catalogue, but not the other.  Figure \ref{fig-complete} shows that our galaxy sample is $>90$\% complete
to $r\approx21$, and falls to 50\% completeness at $r=21.7$.  For our adopted cosmology, 
$r=21.7$ corresponds to an absolute magnitude of $M_r=-17.5+5\log{h}$, including a K-correction of
0.23 mag\footnote{Based on model spectral energy distributions of Coleman, Wu and Weedman (\cite*{CWW}), independent of
SED type to within 0.02 mag.}.  

We used this matched sample to correct the zero points of our $r$ photometry, on a CCD frame by frame
basis, by computing the mean offset $\Delta r$, between the CNOC1 photometry and 
our $r$ magnitudes, for the galaxies we have
in common brighter than $r=21.7$.  We adjusted our $r$ magnitudes according to this difference, which is,
in general, $|\Delta r|<0.3$ mag.
After applying this correction, the difference between the two measures for galaxies brighter
than $r=21.7$ has a 3$\sigma$-clipped {\it r.m.s.} of 0.12 mag (to be compared with 0.21 mag, before
this correction).   This procedure allows us to apply a
reliable magnitude cut to the final sample, by ensuring the $r$ photometry is on a consistent system.
However, the same 
correction cannot be applied to the other filters, as the zero point may vary
between exposures.  In Table \ref{tab-cat}, the original $r$ magnitudes, and their uncertainties,
are given in columns 8 and 9; the corrected magnitude is shown in column 10.

To correct for the incompleteness that sets in around $r$=21.1, we calculated a statistical weight, $R_{\rm weight}$,
which is the ratio of the number of galaxies in the (area matched) CNOC1 photometric sample to the number in the current 
sample, binned in magnitude.  We approximate $R_{\rm weight}=1.12$ for $r<21.1$ and $R_{\rm weight}=1.78r-36.4$ for $21.1<r<22.1$;
this is tabulated in column 17 of Table \ref{tab-cat}.
This weight is applied when considering the statistical properties of the full sample in \S\ref{sec-full}. 
 
There are 165 galaxies in our H$\alpha$ sample which have spectra available from the CNOC1 sample.
For these, the ppp number from the CNOC1 catalogue is given in
column 16 of Table \ref{tab-cat}.  
From the redshifts, we find 136 galaxies for which the H$\alpha$ emission line lies 
within the 10\% transmittance levels of our on-line filter ($0.200<z<0.260$): we classify these as
cluster members, as noted in column 18 of Table \ref{tab-cat}.
From the CNOC1 spectroscopic catalogue, 90.2\% of galaxies in this redshift range lie within
3$\sigma$ of the cluster redshift, and 96.7\% lie within 6$\sigma$, where $\sigma=1095$ km/s is the cluster
velocity dispersion. 

\section{Results} \label{sec-results}
The goal of this project is to detect only the galaxies with the strongest
star formation rates, $\dot{M}\gtrsim 0.5 h^{-2} M_\odot$yr$^{-1}$.  The local analogues
of these galaxies are spiral galaxies of type Sc or later, and starburst galaxies
(e.g. \cite{K92,J+99}).  This should be borne in mind when considering these results,
which do not necessarily apply to the more common types of spiral galaxies.  
\subsection{Comparison with CNOC1 \OII\ Measurements}\label{sec-compareoii} 
For the 165 galaxies with CNOC1 spectra available,
we can compare the \wha\ measurements with their other optical spectral properties.
In particular, we will concern ourselves with the rest frame equivalent widths of the H$\delta$ absorption line, \hd,
and the \OII\ emission line, \ow, as defined in Balogh et al. (\cite*{PSG}).  The former index
is positive for absorption features, while the latter is positive in the case of emission.
The CNOC1 spectroscopic sample is a subset of a complete photometric sample, and statistical weights discussed
in Yee et al. (\cite*{YEC}) are applied to galaxies in this sample to correct for selection effects due to apparent magnitude
($W_m$), geometric position ($W_c$) and colour ($W_c$), where necessary (as in \S\ref{sec-full}). 

\begin{figure*}
\begin{center}
\leavevmode \epsfysize=8cm \epsfbox{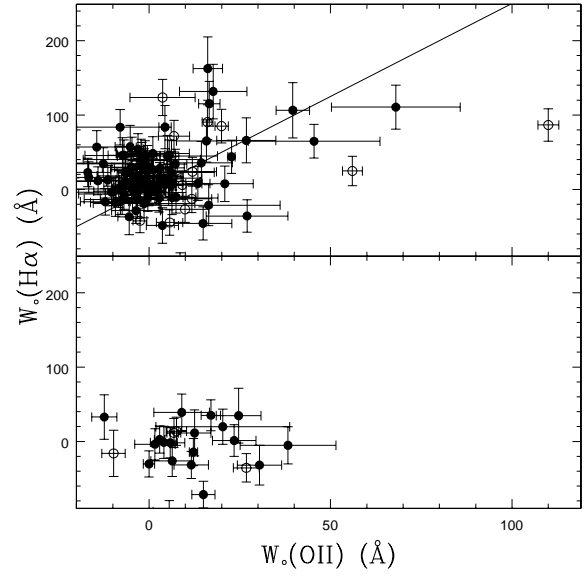}
\end{center}
\caption
{For galaxies with redshifts available from the Abraham et al. (1996) catalogue, \wha\
from our OSIS observations are compared with \ow\ measured from the CNOC1 spectra.
Only galaxies with \ow\ uncertainties less than 20\AA\ are shown.
Field galaxies are plotted in the {\it bottom panel}, and cluster members in the
{\it top panel}.  The {\it solid line} is the mean local relation from Kennicutt (1992).
Deblended galaxies, and those with bright neighbours, are plotted as {\it open symbols}.
Error bars are 1$\sigma$.
\label{fig-oiiha}}
\end{figure*}

In Figure \ref{fig-oiiha} we compare \wha\ with \ow, plotting only galaxies brighter than $r=21.7$, and with
\ow\ uncertainties less than 20\AA.
In the bottom panel we plot field galaxies, for which the H$\alpha$ emission line does not fall within our
on--line filter wavelength range and, thus, should (usually) have \wha=0.  These measurements are fairly evenly distributed about zero (the
median is -1.9\AA), with a standard deviation (excluding 3 outliers with large negative values) 
of $\sim 24$\AA, consistent with the mean uncertainty of $23$\AA.
As expected, we only detect ($>1\sigma$) galaxies with fairly strong emission lines, \wha$\gtrsim 50$\AA.  
In the top panel we show the relation between \ow\ and \wha\ for cluster
members; the solid line represents the mean local relation found by Kennicutt (\cite*{K92}).  Most of the
galaxies that have strong \ow\ (i.e., $\gtrsim 20$\AA) are also detected in H$\alpha$.  Low
ratios of \wha/\ow\ in some cases are partly due to the fact that the H$\alpha$ emission originates from a small (usually
central) region of the galaxy; 
thus, \wha\ is low since there is a considerable
amount of continuum flux dominating the light, from other regions of the galaxy.  The spectroscopic observations,
from which \ow\ is measured, only sample the light in a narrow 1\farcs5 slit (corresponding to
10 pixels in our images, or 3.6 $h^{-1}$ kpc) placed across the galaxy, so the
emission line flux may contribute a larger fraction of the total light in the slit.  

\begin{figure*}
\begin{center}
\leavevmode \epsfysize=8cm \epsfbox{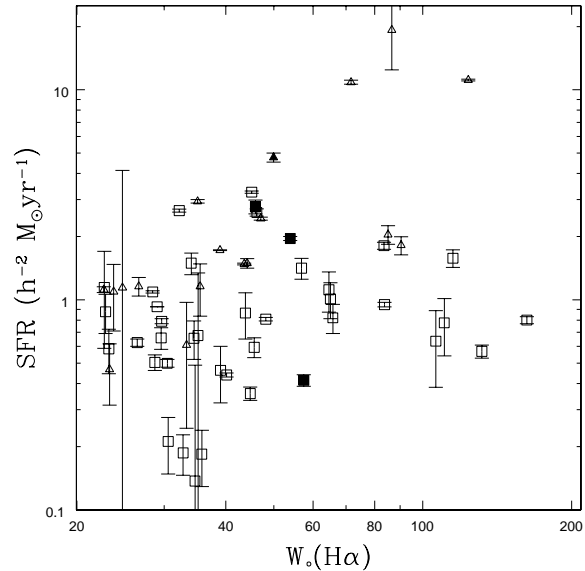}
\end{center}
\caption
{Star formation rates computed from Equation \ref{eqn-sfr}, as a function of \wha, for 
confirmed cluster members brighter than $r=21.7$ and \wha$>20$.  {\it Filled symbols} represent galaxies
for which the H$\alpha$ detection is significant at the 2$\sigma$ level, while
\ow$<5$\AA\ (also $2\sigma$).  Those galaxies plotted as {\it triangles}
were originally blended with another object, or have a nearby, bright neighbour that may
bias the photometry.  Note that the absolute value of the SFRs are systematically
uncertain (by $\sim 30$\%) since the flux calibration is poor; however the relative values should be
reliable.
The SFRs are corrected for [\NII] emission
and 1 magnitude of extinction, but the plotted values of \wha\ are not.
Error bars are only plotted on one axis, for clarity.
\label{fig-sfrha}}
\end{figure*}

We can estimate the SFR
for galaxies detected in H$\alpha$ from Equation \ref{eqn-sfr}, adopting
a mean H$\alpha$ extinction of 1 magnitude and correcting for [\NII] contamination
assuming [\NII]/H$\alpha$=0.3 (\cite{T+99}).  We determine the H$\alpha$ flux
as $f(H\alpha)=f_{on}^{\prime}-f_c$, and the uncertainty in this
quantity includes the 5\% uncertainty in the scale factor $s$.  The luminosity
is computed from the luminosity distance to the cluster of 752 $h^{-1}$ Mpc corresponding
to our adopted cosmology.
The relation between SFR and \wha\ is shown
in Figure \ref{fig-sfrha}; only galaxies brighter than $r=21.7$ with \wha$>20$ \AA\ and
CNOC1 spectra available are
plotted.  Galaxies which were originally blended with another object, or which have nearby, bright
neighbours which may bias the photometry, are plotted as triangles.
Of the 32 cluster members with \ow$<5$\AA\ (at 2$\sigma$ confidence), 
four have \wha$>0$ at the 2$\sigma$ level\footnote{We note that one H$\alpha$-detected galaxy in the catalogue, ppp \#300034,
has an incorrect redshift in the CNOC1 catalogue, and is therefore catalogued with weak [OII].  Using the correct
redshift of 0.22630, the [OII] emission is significant.}; these are shown as the filled symbols 
in Figure \ref{fig-sfrha}.
These four galaxies have 
SFRs of 2--4 $h^{-2} M_\odot \mbox{yr}^{-1}$, comparable to that of the Milky Way
(\cite{Rana}), which suggests that $\sim$12\% of galaxies for which we do not
detect \OII\ may in fact have substantial star formation activity.    
In one galaxy, the H$\alpha$ emission
is confined to a small central region, perhaps indicative of an active nucleus.  Thus, only three of the galaxies 
with undetected \OII\ show significant H$\alpha$ emission extended over large scales;
this comprises only $2\pm1$\% of the galaxies with CNOC1 spectra.
The value of \wha\ for these galaxies is $\sim 50$ \AA; from Kennicutt's (\cite*{K92})
mean relation (see Figure \ref{fig-oiiha}),
this corresponds roughly to \ow$\approx$20 \AA, which should be easily detectable.  
Dust obscuration 
may be responsible for the missing \OII\ in these three galaxies.  However, there is often
a significant amount of emission in the outer, disk regions of these galaxies, which
may have fallen outside the slit in the spectroscopic survey; this aperture effect
may be more important than dust obscuration at 'hiding' signs of star formation, in
some cases.

Interestingly, we also find that 3/8 galaxies which have [OII]$>20$ \AA\ (2$\sigma$) are
undetected in H$\alpha$.  We cannot draw strong conclusions from this, since the uncertainties on
the \wha\ measurements are $\sim 20$\AA, and there could be considerable emission below
our detection limit of 50\AA.  We note that Tresse et al. (\cite*{T+99}) find that $\sim 4$\% of galaxies without
H$\alpha$ emission have detectable [OII] emission.

\subsection{Properties of the Full H$\alpha$ Sample}\label{sec-full}
\begin{figure*}
\begin{center}
\leavevmode \epsfysize=8cm \epsfbox{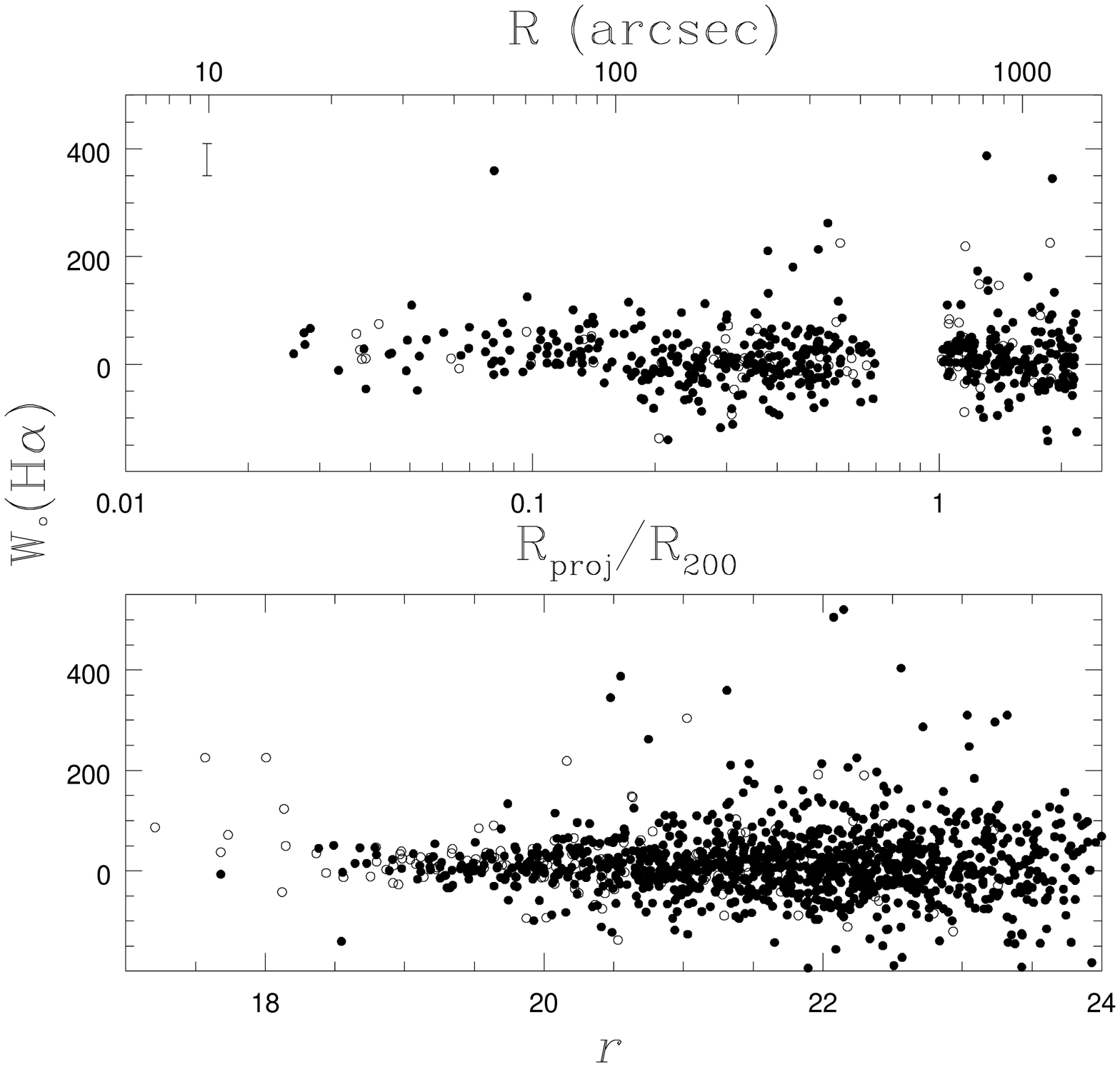}
\end{center}
\caption
{In the {\it bottom panel}, we show \wha\ measurements for all galaxies in the sample, as 
a function of $r$ magnitude.  The sample is 50\% complete at $r\approx 21.7$; galaxies brighter than
this limit are shown in the {\it top panel}, as a function of projected distance from the BCG, normalised
to $R_{200}$ on the bottom scale, and in arcseconds on the top scale.  Open symbols correspond to
galaxies which were deblended, or with nearby, bright neighbours, and for which measurements may be less reliable.
The sample error bar represents the median 1$\sigma$ uncertainty.
Galaxies with H$\alpha$ emission
tend to be fainter than $r=20$, and to be located at large distances from the cluster centre.
\label{fig-Hafig}}
\end{figure*}

\wha\ measurements for the full catalogue are shown in Figure \ref{fig-Hafig}.  In the bottom
panel we show all galaxies, as a function of $r$ magnitude; in the top panel, \wha\ is plotted against
radius, for only those galaxies (581) brighter than our nominal magnitude limit of $r=21.7$.  
Note that both of these figures will include
field galaxies, for which the \wha\ index is not centred on H$\alpha$, and is not correlated
with the strength of the H$\alpha$ line.  There is a clear trend for galaxies with
strong \wha\ to be located at large distances from the cluster centre ($\gtrsim 0.1 R_{200}$).
The single galaxy (apart from the BCG) in the central regions ($R<0.3R_{200}$) with a large H$\alpha$ equivalent width is a
disk galaxy, and the emission seems to be confined to the disk component.  The archival HST image 
shows several bright knots within this disk, and clear signs of interaction with a nearby
galaxy of similar brightness.  The resolution and signal in our H$\alpha$ image is not high
enough to allow a good determination of where the line emission is coming from, however.

\begin{figure*}
\begin{center}
\leavevmode \epsfysize=8cm \epsfbox{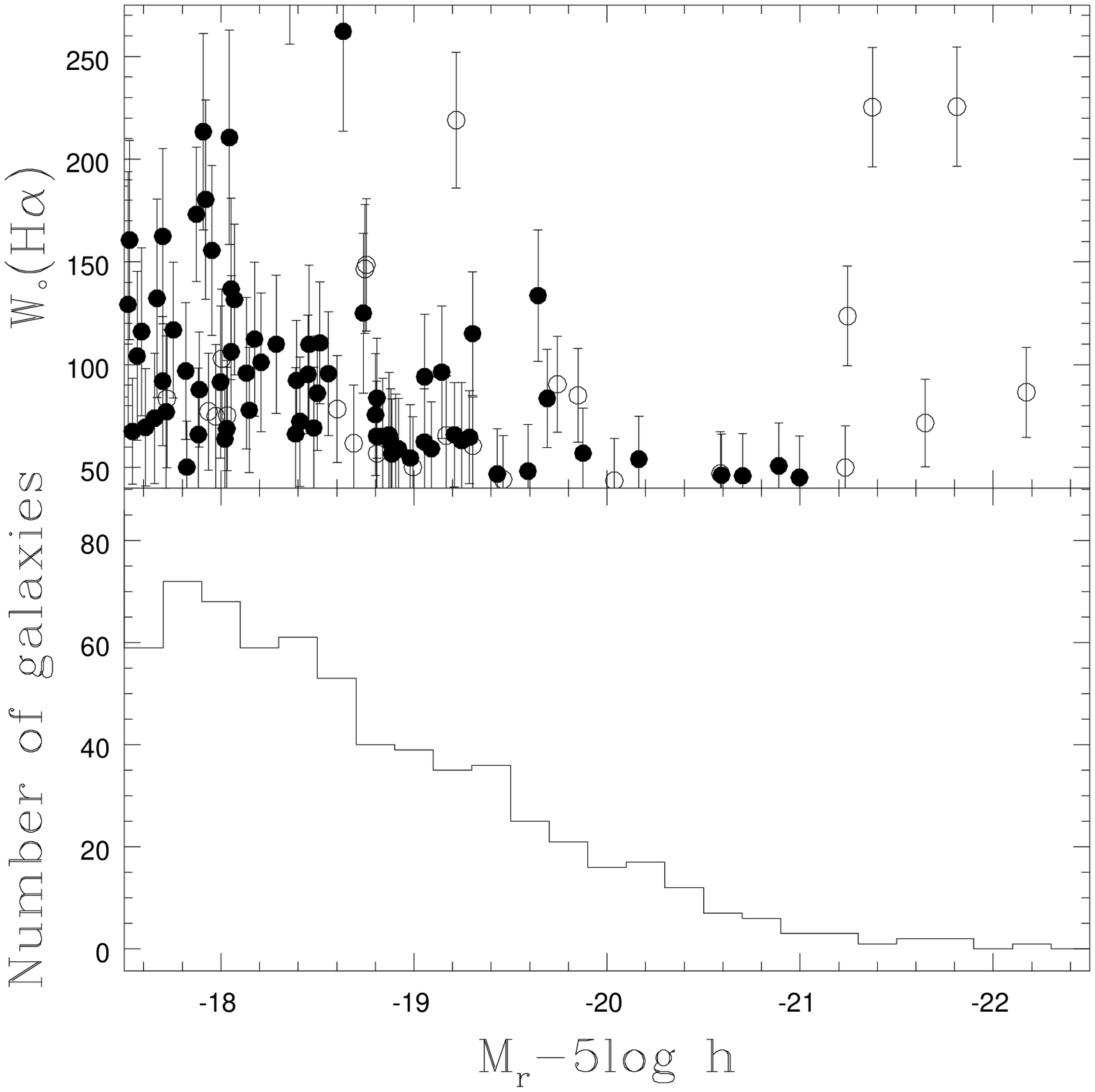}
\end{center}
\caption
{In the {\it top panel}, we show \wha\ measurements, with 1$\sigma$ uncertainties, 
for all galaxies detected in
H$\alpha$ at the 2$\sigma$ level, as a function of their absolute $r$ magnitude.  
Open symbols correspond to galaxies which were deblended,
or with nearby, bright neighbours which may bias the photometry.  The strongest emission lines
are seen in the least luminous galaxies.  For reference, the luminosity function of the
full sample is plotted
in the {\it bottom panel};  since the faintest galaxies are the most numerous, the 
fraction of detected objects is approximately independent of magnitude.
\label{fig-Hamag}}
\end{figure*}

There is also an indication that the strong emission line galaxies tend to be
faint ($r\gtrsim20$).  We demonstrate this further in Figure \ref{fig-Hamag}, where we
compare \wha\ with absolute (K-corrected) $r$ magnitude for all galaxies detected at the 2$\sigma$ level.
Most of the detections are for galaxies with $M_r>-20+5\log h$; and the strongest lines (\wha$>$100 \AA) are
found in galaxies fainter than $M_r=-18+5\log h$.  Since the faintest galaxies are also the most
numerous, as seen in the magnitude distribution of the full sample in the bottom panel, 
the fraction of detections is actually nearly independent of magnitude, at $\sim11$\%.
Despite the large equivalent
widths of these faint galaxies, their corresponding SFRs are fairly low; this can be seen in
Figure \ref{fig-sfrha}, where there is a substantial population of galaxies with \wha$>100$\AA, and
SFR$<$1 $h^{-2}M_\odot yr^{-1}$.  From Jansen et al. (\cite*{J+99}), we expect most of these
faint, strong emission line galaxies to be of Hubble type Sc or later.

\begin{figure*}
\begin{center}
\leavevmode \epsfysize=8cm \epsfbox{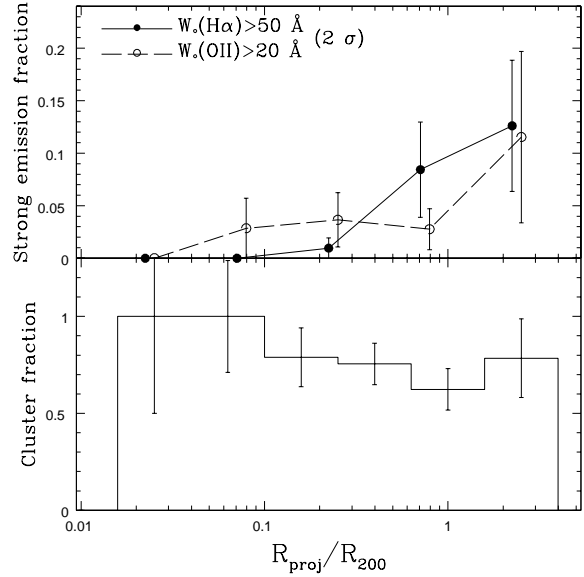}
\end{center}
\caption
{{\it Bottom panel:} The ratio of cluster members to total number of galaxies (brighter than $r=20.7$) 
in the CNOC1 spectroscopic catalogue, weighted
appropriately as discussed in \S\ref{sec-compareoii}, as a function of projected distance from the BCG, normalised
to $R_{200}$.  Error bars are 1$\sigma$, and are computed as the plotted value divided by the square root
of the number of cluster galaxies in that bin.  
{\it Top panel:} The fraction of galaxies with \wha$>50$\AA\ (2$\sigma$ confidence) is plotted as the
solid symbols and solid line; this is statistically corrected for the presence of field galaxies
using the relation in the bottom panel.  From the CNOC1 survey, the fraction of cluster members
with \ow$>20$ \AA\ (again with 2$\sigma$ confidence), which is roughly comparable to the chosen limit
in H$\alpha$, is shown as the open symbols connected by the dashed line (slightly offset in the radial
direction, for clarity).  Error bars are 1$\sigma$.
\label{fig-fHa_new}}
\end{figure*}

To calculate the fraction of cluster galaxies with detected H$\alpha$ emission, we need to make a statistical
correction for the inclusion of field galaxies, which will be relatively more common far from the cluster
centre.  To do this, we use the CNOC1 spectroscopic sample to calculate the fraction of 
of cluster members as a function
of $R_{proj}/R_{200}$, including the appropriate statistical weights discussed in
\ref{sec-compareoii}.  This fraction is shown in the bottom panel of
Figure \ref{fig-fHa_new}; nearly all of the galaxies within 0.1$R_{200}$
are expected to be cluster members, while only $\sim$70\% near $R_{200}$ are members.
Using this statistical field correction, and the $R_{\rm weight}$ weights  discussed in \S\ref{sec-complete}
to correct for incompleteness, we show,
in the top panel of Figure \ref{fig-fHa_new}, the fraction of galaxies that have \wha$>50$ \AA\ with 2$\sigma$ confidence, 
as a function of $R_{proj}/R_{200}$.  We only consider galaxies brighter than $r>20.7$ to allow a fair
comparison with the CNOC1 spectroscopic survey, which is very incomplete (magnitude weights greater
than 5) at fainter magnitudes. 
A total of 4.3$\pm$1\% of all cluster members in the
sample satisfy this criterion, and there is a clear increase in the incidence of this fraction 
with increasing distance from the cluster centre.    
From Kennicutt's (\cite*{K92}) relation, \wha$=50$\AA\ corresponds to \ow$\approx20$\AA. 
The fraction of galaxies in the CNOC1 sample with \ow$>20$\AA, with 2$\sigma$ confidence and to 
the same magnitude limit, is shown in Figure \ref{fig-fHa_new}
as the dashed line; it agrees well with the fraction of strong H$\alpha$ emitting galaxies: within
$R<R_{200}$, the \OII\ fraction is also $4.3\pm1$\%.  
This is a crude comparison, because the detection limit in one
line cannot be uniquely identified with a limit in the other.  However, it is clear that 
the fraction of strong emission line galaxies calculated in the CNOC1 survey (\cite{B+97}), at least for
this cluster, is not strongly underestimated due to aperture or dust effects.

\subsection{H$\alpha$ Emission in Unusual Galaxy Types}
\subsubsection{The cD Galaxy}
The extended distribution of H$\alpha$ emission in the central, cD galaxy of this cluster has
been discussed in Hutchings \& Balogh (\cite*{HB}).  To briefly summarize the results of that paper,
the H$\alpha$ emission extends several arcseconds from the galaxy centre, to the northwest, roughly following
L$\alpha$ and blue knots seen in HST images and spectra.  However, there
is inexact correspondence between H$\alpha$ and L$\alpha$, which may be the result of an 
inhomogeneous dust distribution.  This is the only strong line emitting galaxy (but one) 
detected in the central region
of A2390, and has been omitted from all other analysis in this paper.

\subsubsection{Strong Balmer Line Galaxies}
In the local universe, typical galaxies, regardless of Hubble type, have Balmer absorption lines which are
much weaker than the maximum possible (for a pure A-star composition), due to the
presence of very young or very old stars.  The strongest lines are generally found in
Sc and Sd galaxies, and even in these galaxies \hd$<$5\AA.  Stronger absorption 
seen in a rare class of galaxies seems
to require something unusual in their recent star formation history.  For those strong
Balmer line galaxies
without detectable \OII\ emission lines, it has been suggested that star formation has
recently been truncated, perhaps preceded by a starburst (e.g., \cite{DG83,CS87,Barger,P+99,PSG}).
The remainder (i.e., those with \OII) may be dust-obscured starburst galaxies (\cite{P+99})
in which light from the most
massive stars is suppressed enough to produce the strong Balmer lines observed.

We adopt the definitions of Balogh et al. (\cite*{PSG}) which are \hd$>5$\AA\ and \ow$<5$\AA\
for k+a galaxies, and  \hd$>5$\AA\ and \ow$>5$\AA\ for a+em galaxies.  In the present sample, there are fourteen
galaxies satisfying these constraints, for which we have both CNOC1 spectra and H$\alpha$
measurements.  However, as we wish to consider these objects on a per-galaxy basis, rather
than in a statistical sense, we restrict our sample to those galaxies for which their
spectral classification is the most secure.  We therefore require that the uncertainties
on \ow\ be small enough that the determination of \ow$<5$ \AA\ (for k+a galaxies) or
\ow$>5$ \AA\ (a+em galaxies) is significant at the 1$\sigma$ level.  We also remove one
galaxy with a large uncertainty in \hd, and include another which lies just below our \hd\
threshold (4.7\AA), but has \hd$>3$ \AA\ with more than 2$\sigma$ confidence.
We list the properties of these selected galaxies in Table \ref{tab-kaak}. 

\begin{deluxetable}{cccccccl}
\tablewidth{0pt}
\footnotesize
\tablecaption{Strong Balmer Line Galaxies\label{tab-kaak}}
\tablehead{
\colhead{ppp}&\colhead{\ow}&\colhead{$\Delta$\ow}&\colhead{\whd}&\colhead{$\Delta$\whd}&\colhead{\wha}&\colhead{$\Delta$ \wha}&\colhead{Comments}
}
\startdata     
101695 & 16.6 &  2.9 & 5.8 & 1.4 & 115.1 & 30.2& a+em: central emission\nl 
500858 & 14.4 &  8.5 & 8.1 & 4.3 & 35.8 & 28.7&  a+em\nl
101100 & -1.0 & 3.1 &  4.7 & 0.7 & 45.1 & 20.2& (almost) k+a: diffuse emission\nl
400656 &  1.2 & 1.6 &  5.3 & 1.1 & 26.5 & 20.4& k+a: bow shock?\nl
100537 &  0.0 & 2.7 &  5.5 & 0.6 & -2.8 & 18.0& k+a\nl
201228 & -7.3 & 10.1&  5.8 & 2.3 & 45.6 & 23.7& k+a: emission in central knot\nl
100604 & -9.3 & 6.9 &  5.9 & 1.6 & -17.6 & 18.8& k+a\nl          
501033 & -8.0 & 11.9&  7.1 & 2.1 & 83.5 & 23.8& k+a: emission in disk\nl
\enddata
\tablecomments{All equivalent widths are rest frame, in units of \AA.  Uncertainties are 1$\sigma$.}
\end{deluxetable}

\begin{figure*}
\begin{center}
\leavevmode \epsfysize=8cm \epsfbox{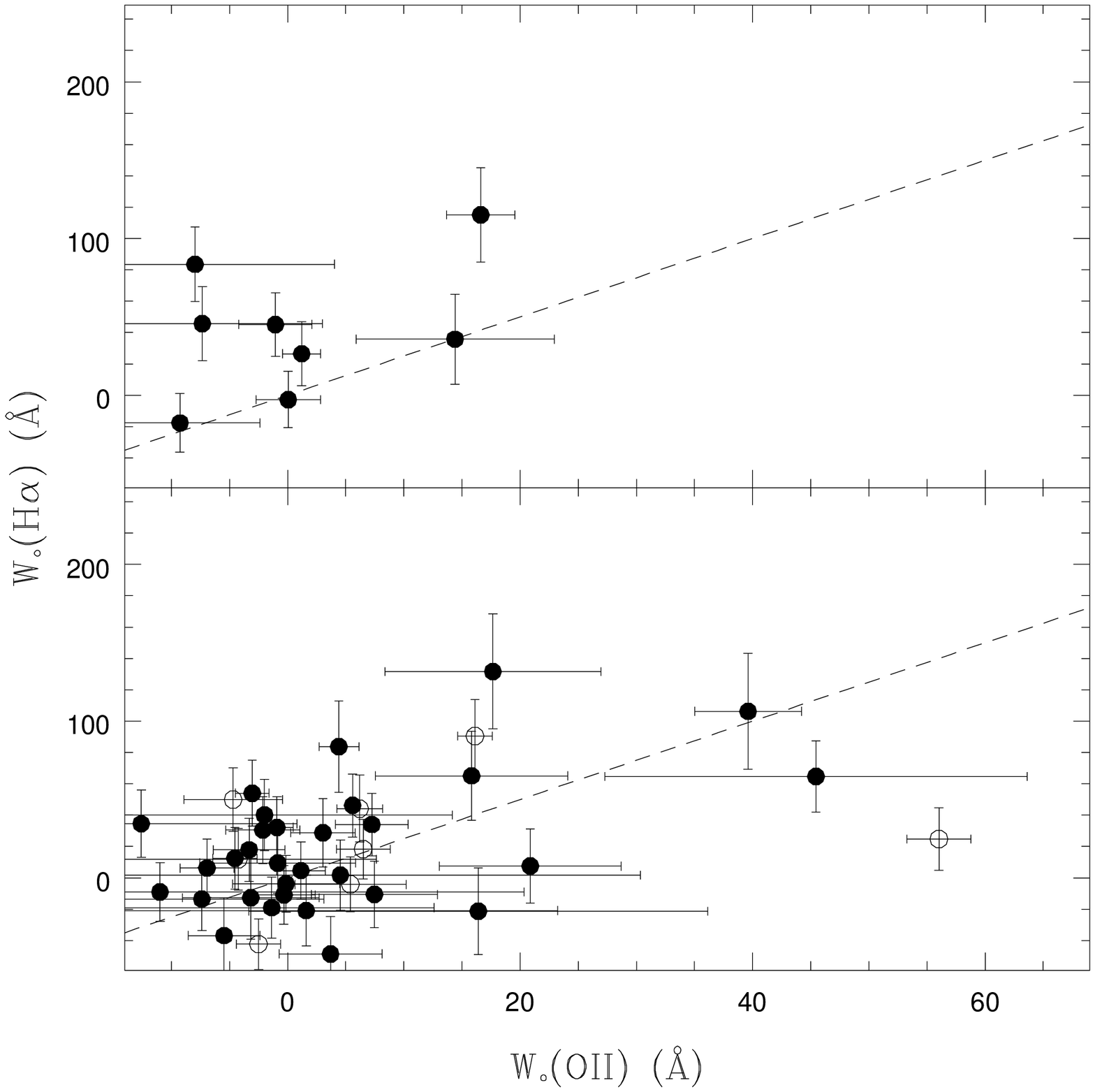}
\end{center}
\caption
{We show the \wha--\ow\ relation for galaxies matched with the CNOC1 spectroscopic sample,
with uncertainties on \ow\ less than 40\AA.  Error bars are 1$\sigma$. 
In the {\it bottom panel} we show galaxies with $(g-r)<0.8$ and \hd$<3$ \AA, which represents
the ``normal'' population of blue galaxies.  In the {\it top panel} we show
those galaxies that we have defined as secure k+a or a+em galaxies (see text for definitions).
The {\it dashed line} in both panels represents the local relation of Kennicutt (1992).
{\it Open symbols} identify those galaxies which were originally blended with another, or which have
bright neighbours which may bias the photometry.
\label{fig-kaak}}
\end{figure*}

In the top panel of figure \ref{fig-kaak} we show the \ow-\wha\ relation for 
these eight galaxies.  
For comparison, we present the distribution of
blue ($(g-r)<0.8$) galaxies with \hd$<3$ \AA\ in the bottom panel.
The eight H$\delta$-strong galaxies appear to lie toward somewhat higher \wha/\ow\ ratios;
however, there are not
enough objects to allow strong conclusions.  Below, we discuss the H$\alpha$ and continuum
morphologies of the k+a and a+em galaxies separately, beginning with the four k+a galaxies
with \wha\ $>0$.

\begin{figure*}
\begin{center}
\leavevmode \epsfysize=12cm \epsfbox{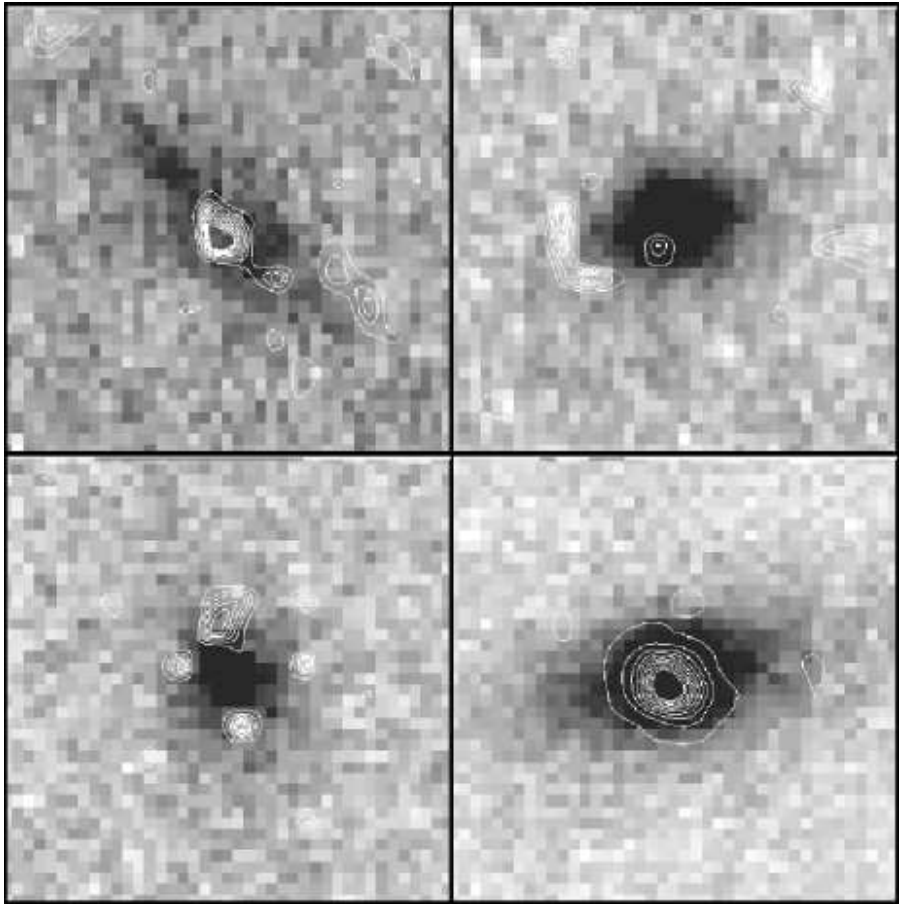}
\end{center}
\caption
{Three k+a galaxies and one a+em galaxy with detected H$\alpha$ are shown.  The galaxies
show no signs of interaction in the continuum light (greyscale image).  The H$\alpha$
contours are overplotted in logarithmic intervals, after smoothing with a 3 pixel boxcar filter.
Each panel is 6\farcs25 wide.  Northwest
is to the bottom left.  The four galaxies show a variety of 
emission line morphologies, in particular (from the top left, clockwise): 1) a+em: central, extended
emission in a disk galaxy; 2) k+a: an arc of emission, possibly indicating a shock front; 
3) k+a: diffuse, extended emission over large scales; and 4) k+a: emission, possibly from HII regions in a disk.
\label{fig-mosaic}}
\end{figure*}

\begin{itemize}
\item[501033:] The only k+a galaxy for which \wha$>0$ with more than 3$\sigma$ significance,
this object is shown in the bottom left panel of Figure \ref{fig-mosaic}.  The  emission comes from 
the disk (perhaps from individual \HII\ regions, though this conclusion depends on the
smoothing scale adopted), with no central emission.   
The absence of \OII\ in the spectrum may be partly due to an aperture effect, since some of the
disk emission will not have been covered by the 1\farcs5 slit (10 pixels in our images).  However, it seems unlikely
that this can be the only reason, as most of the disk should still have been covered.
\item[101100:] This galaxy is detected in H$\alpha$ at the 2.2$\sigma$ level, and
is shown in the bottom
right panel of Figure \ref{fig-mosaic}.  Although the value of \hd\ falls just below our adopted
threshold of 5\AA\ for k+a classification, the H$\delta$ line is clearly strong, greater than
3\AA\ with more than 2$\sigma$ significance.
Unlike galaxy 501033, the H$\alpha$ emission
in this object is weakly present over large scales, and is peaked on the nucleus.  In this
case, the \OII\ light
must be obscured, probably by dust in the centre of the galaxy. 
\item[201228:]This is not quite a 2$\sigma$ detection in H$\alpha$.
The emission in this k+a galaxy arises from a compact region at the centre, 
which may indicate the presence of a non-thermal nuclear source. 
\item[400656:] The final k+a galaxy detected in H$\alpha$ is shown in the top right panel of
Figure \ref{fig-mosaic}.  The emission forms a distinct arc on the
western edge.  Although the formal significance of the detection is low, this is
because the emission arises from a much smaller area than the bulk of the continuum light.
At the position of the arc, the H$\alpha$ flux is clearly detected at $>2\sigma$.
The centre of the cluster is almost due west from this bulge-dominated galaxy; thus, it is possible
that this emission arises from a shock, or induced star formation,
 due to an interaction with the intercluster medium as the galaxy plunges toward the
cluster centre.  However, the image quality is not good enough to rule out the possibility
that the emission shape arises normally in a spiral arm, or in a tidal tail.  Since the long edge
of the CNOC1 slit was aligned in the east--west direction, this emitting region should have 
contributed to the spectrum obtained in that survey.
\end{itemize}
In summary,
only 2/6 of the k+a galaxies are undetected in H$\alpha$ as well as \OII\ (though the
formal significance of two others is less than 2$\sigma$ due to the relatively small area
in which the emission originates). In one case 
the absence of \OII\ may be partly attributed to an aperture effect.  In two other galaxies,
the H$\alpha$ emission may arise from non--thermal sources (i.e. active nucleus or shocked gas).  Therefore, there are only
two k+a galaxies in which there is strong H$\alpha$ resulting from star formation in the regions
covered by the spectroscopic slit; it is possible that dust obscuration is responsible for the
absence of \OII\ in the CNOC1 spectra of these objects.  It seems probable that there
is substantial star formation taking place in at least some galaxies that were classified as
k+a based on spectra that did not include H$\alpha$ emission.  We can therefore conclude that
galaxies with truncated star formation likely make up only a subset of the k+a class of galaxies.

We now consider the a+em galaxies, in which the H$\delta$
lines are too strong to be consistent with the ongoing star formation implied by the
presence of \OII\ emission (\cite{P+99,PSG}).  Our sample only
contains two such galaxies, and only one is
detected in H$\alpha$ at the $2\sigma$ level.  The detected galaxy  is 
shown in the top, left panel of Figure \ref{fig-mosaic}:
it is a pure disk galaxy, with fairly symmetric, diffuse emission centred on the continuum peak.
The ratio \wha/\ow\ $=6.9\pm2.8$, which is larger than the mean ratio found by Kennicutt (\cite*{K92})
by 1.6$\sigma$, as expected if there is considerable dust extinction.  However, we cannot draw any
conclusions from this one object.  The only other secure a+em galaxy has a ``normal'' ratio,
with fairly large uncertainties on both line indices. 

\section{Conclusions} \label{sec-concs}
We have presented \wha\ measurements for the strongest emission line cluster galaxies (corresponding to
late type spirals and starburst galaxies) from a sample of 1189 galaxies over fields covering 270 \sq\arcmin\ about the cluster
Abell 2390. We confirm the presence of a gradient in strong emission line frequency within this cluster;
the fraction of galaxies in which H$\alpha$ emission is detected at the 2$\sigma$ level increases from 0.0 in the central 
regions ($R_{proj}/R_{200}<0.02$, excluding the BCG) to $\sim 10$\% at $R_{200}$.   This is consistent
with the fraction of galaxies with strong \OII\ emission, measured in the CNOC1 survey. 
We compare \wha\ with \ow\ measured from CNOC1
spectra for 166 galaxies, and find that less than $\sim 3$\% 
show no significant \OII\ emission in their spectra, and yet 
are detected in H$\alpha$ at the 2$\sigma$ level.
The \OII\ emission in these galaxies is likely absent in the spectra due to a variety of effects, such as
1) non-central (i.e., disk) emission missed by the narrow slit; 
2) naturally low \ow/\wha\ ratios due to non-thermal ionization; and 3) dust extinction.
At least two out of the six clear examples of k+a galaxies (which have
strong H$\delta$ absorption but no detectable \OII) have strong, diffuse H$\alpha$
emission within the region covered by the spectroscopic slit; thus, the
fraction of galaxies in which star formation was recently truncated 
is less than estimated from the frequency of these spectral types.

Moss \& Whittle (\cite*{MW00})
have found evidence that circumnuclear star formation in spiral galaxies is more common
in rich clusters than in the field,  and claim this enhanced star formation is due to
cluster tidal effects.  We note that the circumnuclear
emission detected by these authors has, in general, equivalent widths
\wha$<50$\AA\ (\cite{MW}),
and would not have been detected in the present study.  
The lack of dramatically starbursting cluster galaxies in Abell 2390
found in the present survey is consistent with the 
conclusions based on the spectral analysis of this cluster (\cite{A2390,PSG}), that
cluster--induced star formation is unlikely to play a large role in
cluster galaxy evolution.  The level
of star formation observed in this cluster is consistent with the ``strangulation'' model
of cluster galaxy evolution, in which star formation is assumed to decline gradually after
a galaxy is accreted into a cluster (\cite{infall}).  

\acknowledgments 
This work was completed while MLB was supported by a Natural Sciences and Engineering 
Research Council of Canada (NSERC)
research grant to C. J. Pritchet and an NSERC postgraduate scholarship. 
It was completed under support from a PPARC rolling grant for extragalactic astronomy
and cosmology at Durham.  We would like to thank the anonymous referee for many useful comments
which improved this paper.

\bibliographystyle{astron_mlb}
\bibliography{ms_pp}

\end{document}